\documentstyle[11pt,paspconf]{article}
\def\simless{\mathbin{\lower 3pt\hbox
   {$\rlap{\raise 5pt\hbox{$\char'074$}}\mathchar"7218$}}}   
\def\simgreat{\mathbin{\lower 3pt\hbox
   {$\rlap{\raise 5pt\hbox{$\char'076$}}\mathchar"7218$}}}   
\def\etal{{\rm et al.}}
\input psfig.tex

\def\tc {t_{\rm cross}}
\def\tdiss {t_{\rm diss}}

\def\vdsp {v_{\rm disp}}

\def\solmas{{M$_\odot$}}
\def\solm{{M_\odot}}
\def\be{\begin{equation}}
\def\ee{\end{equation}}

\begin{document}

\title{Early Evolution of Stellar Clusters}
\author{Ian A. Bonnell}
\affil{Institute of Astronomy, Madingley Road, Cambridge, CB3 0HA, UK}

\begin{abstract}
Observations have revealed that most stars are born in
clusters.  These systems, containing from tens to thousands
of stars and typically significant mass in gas in the youngest systems,
evolve due to a combination of stellar and star-gas interactions. 
Simulations of pure stellar systems are used to investigate possible
initial configurations including ellipticity, substructure and
mass segregation. Simulations of gas-rich clusters investigate the
effects of accretion on the cluster dynamics and on the individual masses that
result in a stellar mass spectrum. Further stellar interactions,
including binary destruction and eventually cluster dissolution are
also discussed.

\end{abstract}

\keywords{Young Clusters, Dynamics, Star formation, Mass segregation, Stellar masses, Binaries}

\section{Introduction}

Although stellar clusters and associations contain but a small
fraction of the stellar content of the Galaxy, it is becoming
increasingly clear that most stars originate in clusters and that to
understand how stars form we have to consider clusters of stars and
how such environments can affect their formation. Surveys of nearby
star forming regions have found that the majority of pre-main sequence
stars are found in clusters (e.g Lada et. al. 1991; Lada, Strom \&
Myers~1993; see also Clarke, Bonnell \& Hillenbrand~2000).  The
fraction of stars in clusters depends on the molecular cloud
considered but generally varies from 50 to $\simgreat 90$ per
cent. This fraction appears to decreases with age.

Young stellar clusters in the solar neighbourhood are found to contain
anywhere from tens to thousands of stars with typical numbers of
around a hundred (Lada et. al.~1991; Phelps \& Lada~1997; Clarke
et. al.~2000). Cluster radii are generally a few tenths of a
parsec such that mean stellar densities are of the order of $\approx
10^3$ stars/pc$^3$ (c.f. Clarke et. al.~2000) with central stellar
densities of the larger clusters (ie the ONC) being $\simgreat 10^4$
stars/pc$^3$ (McCaughrean \& Stauffer~1994; Hillenbrand \&
Hartmann~1998; Carpenter et. al.~1997).

These clusters are usually associated with massive clumps of molecular
gas. Indeed, the mass in gas is typically much more than that in
stars.  Gas can thus play an important role in the dynamics of the
clusters and possibly affect the final stellar masses through
accretion. 

Recently, it has become possible to conduct a stellar census in young
stellar clusters (e.g. Hillenbrand~1997) by using theoretical pre-main
sequence evolutionary tracks to estimate each star's mass and
age. Unfortunately, there is a fair degree of uncertainty in these
estimates due to the uncertainty in the tracks themselves (see
Hillenbrand~1997 for an example) and due to the possibility of ongoing
gas accretion affecting the star's pre-main sequence evolution (Tout,
Livio \& Bonnell~1999). What is relatively certain in the census is
that the cluster stars are generally young (ages $\approx 10^6$ years) and
contain both low-mass and high-mass stars in proportion as you would
expect from a field-star IMF (Hillenbrand~1997). Furthermore, there is a
degree of mass segregation present in the clusters with the most massive stars
generally found in the cluster cores. 

In this paper, I review work that has been done on the
evolution of young stellar clusters and how this work can be combined
with observations of such clusters to investigate the relevant processes
of star formation in clusters.

\section{Cluster Morphology}

\begin{figure}
\centering
\vspace{-0.75truein}
\centerline{\psfig{figure=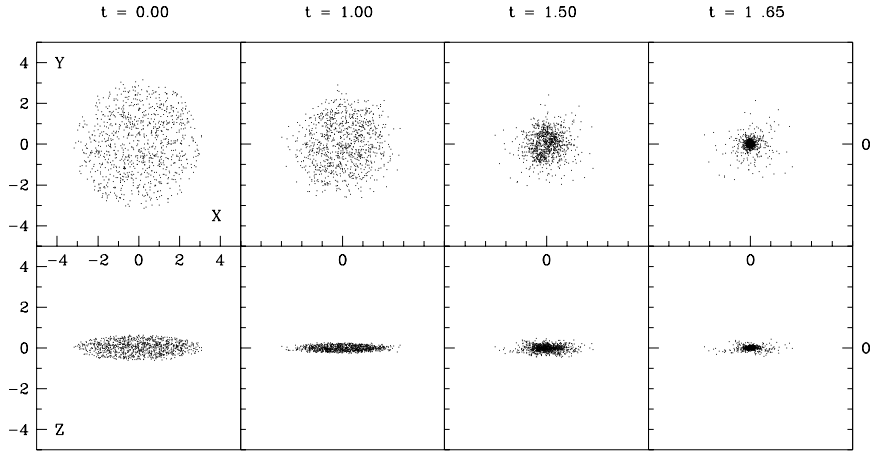,width=5.0truein,height=5.0truein,rwidth=5.0truein,rheight=3.0truein,angle=-90}}
\caption{\label{collrelax} The cold collapse and violent relaxation of an
initially flattened cluster (Boily \etal~1999). The post violent-relaxation
cluster remains flattened, although less so than the original cluster.}
\end{figure}

One of the major questions concerning stellar clusters is their
formation mechanism. A number of different scenarios can be imagined
including a triggering where a shock induces star formation in a
group, or a coagulation where a number of smaller groups merge to form
one cluster (e.g. Klessen, Burkert \& Bate~1998).  Such processes can
leave traces of their initial conditions in the cluster morphologies.
For example, a triggered formation mechanism such as a shock or
cloud-cloud collision leads to a flattened structure which then
fragments to form a cluster (Whitworth \etal~1994, Bhattal \etal~1998).
This flattened morphology should then be in the initial conditions of
the stellar cluster.

In order to use the morphology in present-day clusters to constrain
possible initial conditions, we have to understand how the early
evolution of the stellar cluster can affect its morphology.  Boily,
Clarke \& Murray (1999; see also Goodwin~1997a) have investigated the
evolution of a dynamically cold flattened stellar system. Using N-body
simulations, they  showed how the cluster relaxes through both an
initial violent relaxation and subsequent two-body relaxation. As
these systems are generally younger than their relaxation time (see
below), it is the violent relaxation which will have a greater affect
on the cluster morphology.  Boily \etal~found that a flattened system
becomes less elliptical, but that the violent relaxation does not
completely remove the initial asphericity. For example, a system with
an initial axis ratio of 5:1 relaxed to an axis ratio of 2:1. This
suggests that the clusters such as the ONC which are significantly
elongated (axis ratio of 2:1, Hillenbrand \& Hartmann~1998; see also
Clarke \etal~2000) were initially very aspherical and thus could have
been formed as the result of a triggering mechanism.

\begin{figure}
\centering
\centerline{\psfig{figure=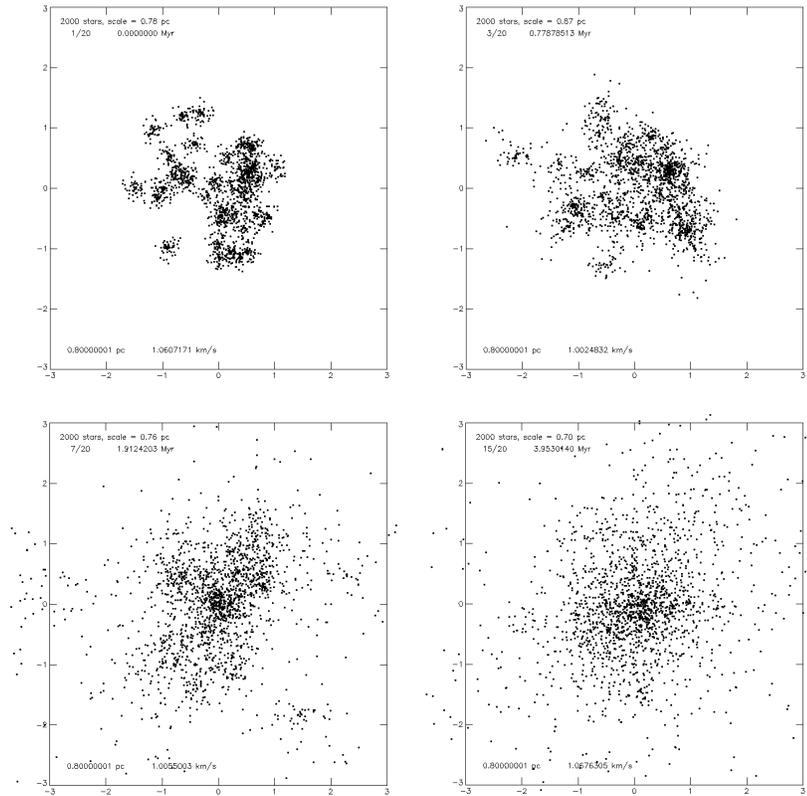,width=4.25truein,height=4.25truein}}
\caption{\label{subclust} The evolution of a cluster containing
subclusters (Scally~in preparation). The subclusters dissolve over time into
the larger-scale cluster.}
\end{figure}

Substructure, or subclustering, is another possibility in the
cluster's initial conditions that can constrain the formation
mechanism.  Bate, Clarke \& McCaughrean~(1998) used statistical tests
for substructure in clusters based on the 
mean surface density of companions (see also Larson~1995). They found
that the ONC is consistent with having no substructure although a narrow
window of subclustering is possible. In contrast, IC348 does appear,
at least by eye, to have significant substructure (Lada \& Lada~1995),
but this has not been tested statistically.

At present, most young clusters do not display significant
substructure, but it is possible that the initial conditions did
contain subclustering that has sinced been removed through its
evolution (Bate \etal~1998). This can be investigated through N-Body
simulations (Scally, in preparation), investigating how the
subclusters relax and dissolve into the surrounding larger-scale
cluster. This occurs due to a combination of the tidal forces acting
on the subcluster and its internal relaxation which leads it to
dissolve on a timescale proportional to the number of stars it
contains.

\section{Mass Segregation}

\begin{figure}
\centering
\vspace{-0.1truein}
\centerline{\psfig{figure=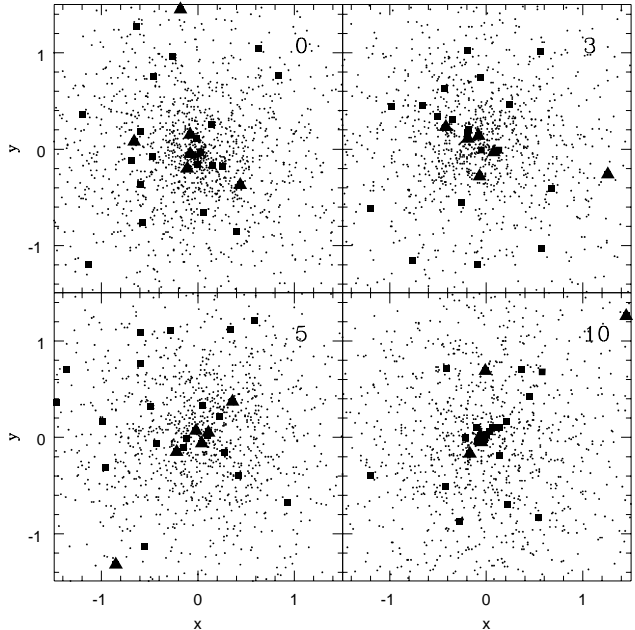,width=3.6truein,height=3.6truein,rwidth=3.6truein,rheight=3.4truein}}
\caption{\label{massseg} The dynamical mass segregation of the massive
stars in a cluster containing 1500 stars (Bonnell \& Davies~1998). The
six most massive stars are indicated by triangles while the next 24
most massive stars are indicated by squares. The time is given in
units of the crossing time.}
\end{figure}

Young stellar clusters are commonly found to have their most massive stars
in or near the centre (Hillenbrand~1997; Carpenter \etal~1997). 
This mass segregation is similar to that found in older clusters but
the young dynamical age of these systems offers the chance to test
whether the mass segregation is an initial condition or due to the
subsequent evolution.
We know that two-body relaxation drives a stellar system towards
equipartition of kinetic energy and thus towards mass segregation.  In
gravitational interactions, the massive stars tend to lose some of
their kinetic energies to lower-mass stars and thus sink to the centre
of the cluster (see Fig.~\ref{massseg}). 

Numerical simulations of  two-body relaxation have shown
that while some degree of mass segregation can occur over the short lifetimes
of these young clusters, it is not sufficient to explain the observations
(Bonnell \& Davies~1998).
Thus the observed positions of the massive stars near the centre of clusters like the ONC reflects the initial conditions of the cluster and of massive star
formation that occurs preferentially in the centre of rich clusters.

Forming massive stars in the centre of clusters is not straightforward
due to the high stellar density. For a star to fragment out of the
general cloud requires that the Jeans radius, the minimum radius for the
fragment to be gravitationally bound, 
\begin{equation} 
R_J \propto  T^{1/2} \rho^{-1/2}, 
\end{equation}
be less than the stellar
separation. This implies that the gas density has to be high,
as you would expect at the centre of the cluster potential. The difficulty 
arises in that the high gas density implies that the fragment mass, being
approximately the Jeans mass, 
\begin{equation}
\label{Jeans_mass}
M_J \propto T^{3/2} \rho^{-1/2}, 
\end{equation} 
is quite low. Thus, unless the
temperature is unreasonably high in the centre of the cluster {\bf
before} fragmentation, the initial stellar mass is quite low.
Equation~(2) implies that the stars in the centre of the cluster
should have the lowest masses, in direct contradiction with the
observations. Therefore, we need a better explanation for the
origin of massive stars in the centre of clusters.

\section{Accretion and stellar masses}

\begin{figure}
\vspace{-0.05truein}
\centerline{\psfig{{figure=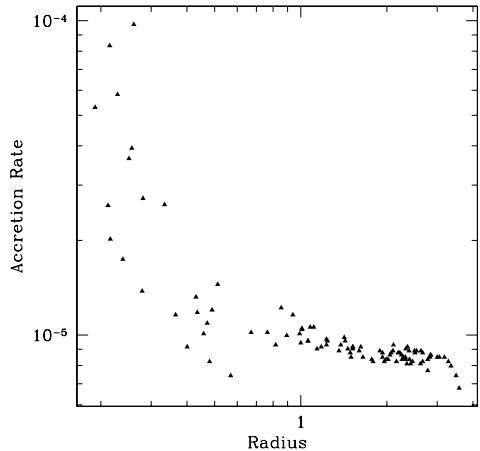,width=2.750truein,height=2.750truein,rwidth=2.75truein,rheight=2.5truein}}}
\caption{\label{cluacc100} The accretion rates versus radius for a cluster
containing 100 stars and 90 \% of its mass in gas.}
\end{figure}

Young stellar clusters are commonly found to be gas-rich with
typically 50 \% to 90 \% of their total mass in the form
of gas (e.g. Lada~1991). This
gas can interact with, and be accreted by, the stars as 
both move in the cluster. If significant accretion occurs, it can affect
both the dynamics and the masses of the individual stars (e.g. Larson~1992). 

Simulations of accretion in clusters using a combination SPH and
N-body code have found that accretion is a highly non-uniform process
where a few stars accrete significantly more than the rest (Bonnell
\etal~1997).  Individual stars' accretion rates depend largely on
their position in the cluster (see Fig.~\ref{cluacc100}) with those in
the centre accreting more gas than those near the outside.  This
process is termed ``competitive accretion'' (Zinnecker~1982) each
star competes for the available gas reservoir with the advantage going to
those in the cluster centre that benefit from the overall cluster potential.

Accretion in stellar clusters naturally leads to both a mass spectrum
and mass segregation. Even from initially equal stellar masses, the
competitive accretion results in a wide range of masses with the most
massive stars located in or near the centre of the
cluster. Furthermore, if the initial gas mass-fraction in clusters is
generally equal, then larger clusters will produce higher-mass stars
and a larger range of stellar masses as the competitive accretion
process will have more gas to feed the few stars that accrete the most
gas.

\begin{figure}
\psfig{{figure=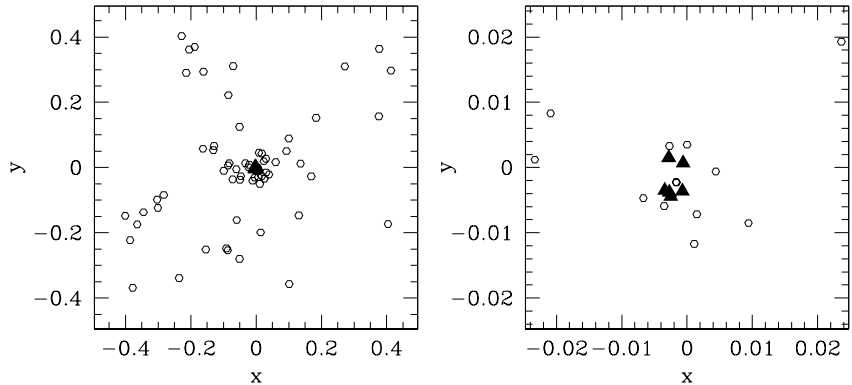,width=4.7truein,height=4.7truein,rwidth=4.7truein,rheight=2.3truein}}
\caption{\label{accmassseg} The position of the 6 most massive stars
(with $M_{\star} \simgreat 4 M_{\rm med}$, filled triangles) in a cluster of
100 stars. The right-hand panel is a blow-up of the cluster core.
The masses of the most massive stars are due to the
competitive accretion process.}
\end{figure}

\section{Formation of Massive Stars}

The formation of massive stars is problematic not only for their
special location in the cluster centre, but also due to the fact that
the radiation pressure from massive stars is sufficient to halt the
infall and accretion (Yorke \& Krugel~1977; Yorke~1993). This occurs
for stars of mass $\simgreat 10$ \solmas. 

A secondary effect of accretion in clusters is that it can force it to
contract significantly. The added mass increases the binding energy of
the cluster while accretion of basically zero momentum matter will
remove kinetic energy.  If the core is sufficiently small that its
crossing time is relatively short compared to the accretion timescale,
then the core, initially at $n\approx 10^4$ stars pc$^{-3}$, can
contract to the point where, at $n\approx 10^8$ stars pc$^{-3}$,
stellar collisions are significant (Bonnell, Bate \& Zinnecker~1998).
Collisions between intermediate mass stars ($2 \solm \simless m
\simless 10 \solm$), whose mass has been accumulated through accretion
in the cluster core, can then result in the formation of massive ($m
\simgreat 50 \solm$) stars. This model for the formation of massive
stars predicts that the massive stars have to be significantly younger
than the mean stellar age due to the time required for the core to
contract (Bonnell \etal~1998).

\section{Binaries in clusters}

\begin{figure}
\centering
\vspace{-0.1truein}
\centerline{\psfig{figure=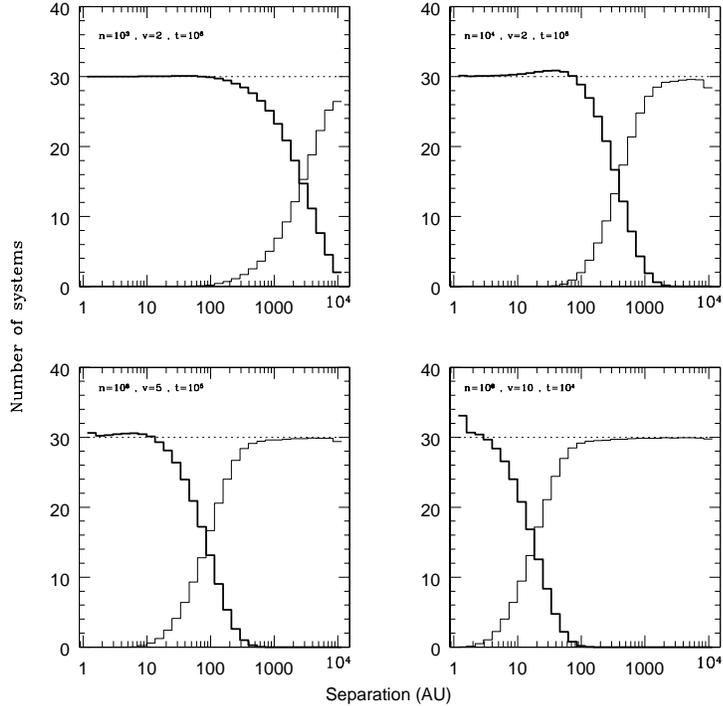,width=4.truein,height=4.truein,rwidth=4.truein,rheight=4.truein}}
\caption{\label{binclus} Binary frequency (heavy line) versus log
separation (in au) for clusters with a stellar density of $n = 10^3$
stars pc$^{-3}$, velocity disperion of $\vdsp = 2$ km s$^{-1}$ for
$10^6$ years (top left panel), $n =10^4$ stars pc$^{-3}$, $\vdsp = 2$
km s$^{-1}$ for $10^6$ years (top right panel),$n =10^6$ stars pc$^{-3}$,
$\vdsp = 5$ km s$^{-1}$ for $10^5$ years (bottom left panel) and $n =10^8$
stars pc$^{-3}$, $\vdsp = 10$ km s$^{-1}$ for $10^4$ years (bottom
right panel). The binary frequency is initially flat and the light
line indicates the number of systems destroyed versus separation (Smith
\& Bonnell, in preparation.}
\end{figure}

Binary stars can play an important role in young stellar clusters as
well as in older, evolved clusters. Their importance stems not from
their influence on the dynamics of the cluster, but rather from how
the cluster dynamics affect, and destroys, some of the binaries.  The
high binary frequency in nearby, non-clustered star-forming regions
(e.g. Mathieu~1994; Ghez~1995, Mathieu \etal~2000) compared to the
main sequence (Duquennoy \& Mayor~1991) and to that in clustered star
forming regions (Padgett, Strom \& Ghez~1997; Petr \etal~1998; Mathieu
\etal~2000) suggest that as most stars are formed in clusters, the
cluster environment plays an important role in setting the binary
frequency. This can happen in one of two ways, either the cluster
environment impedes binary formation or it subsequently destroys them.

Kroupa~(1995) has shown how a 100 \% binary frequency can be reduced through
stellar encounters in clusters and how the final binary frequency depends
on the stellar density in the cluster. Binary systems wider than the
hard-soft limit (basically where the orbital velocity is equal to the
cluster velocity dispersion) are destroyed in encounters. Thus
denser systems with higher velocity dispersions disrupt more binaries.

This binary destruction in clusters can also be used as a tracer of
the cluster evolution (Smith \& Bonnell, in
preparation). Figure~\ref{binclus} shows the resultant binary
frequency versus separation for clusters of various stellar
densities. Clusters with stellar densities of the order of $10^3$
stars pc$^{-3}$ only have a significant effect on systems wider than
$\simgreat 10^3$ au whereas clusters with higher densities destroy
closer systems. Thus, if a cluster does go through a very dense phase
in order for collisions to occur (see above) then no binaries wider
than 100 au should survive with significant depletion extending down
to 10 au (Fig.~\ref{binclus}).

\section{Cluster Dissolution}

The majority of young stellar clusters dissolve before their
low-mass stars reach the main sequence. This can happen either through 
a sudden removal of the majority of the binding mass, the gas contained
in the cluster, or through the dynamical interactions that  put all
of the cluster's binding energy into a central binary. 

Gas removal is most important for large stellar clusters that are more
likely to contain massive stars. These stars can ionise the
intracluster gas which can then escape (unless the velocity dispersion
is $\vdsp \simgreat 10$ km s$^{-}$). Gas removal on timescales less
than the dynamical time is catastrophic for the cluster if gas is the
major mass component (Lada, Margulis \& Dearborn~1984;
Goodwin~1997b). Gas removal over many dynamical times will leave a
remnant cluster containing a fraction of the initial cluster stars.

The second possibility for cluster dissolution is that two-body
relaxation takes the cluster's binding energy and puts it into one
central binary, typically containing the most massive stars (Sterzik
\& Durisen~1998; Bonnell~in preparation). This occurs on a timescale
similar to the relaxation time of the cluster as a whole (the
difference in that it involves only the central binary as the energy
source and that the binary shrinks during the energy exchange),
\begin{equation}
\tdiss \approx N \tc.
\end{equation}
Thus, small clusters will dissolve readily through two-body relaxation whilst
large clusters dissolve through the interaction of their massive stars with the
gas. It should only therefore be the intermediate clusters which have long dissolution times but that do not contain very massive stars which survive long
enough to be considered as open or Galactic clusters.

\section{Summary}

The early evolution of stellar clusters involves many interactions
which affect the clusters' and the individual stellar properties. 
Understanding these interactions, and their possible consequences,
allows us to investigate probable cluster initial conditions
and how they relate to observations of young stellar clusters.

The dynamical interactions are of two types, pure stellar interactions
and star-gas interactions. The first type are investigated through N-Body simulations and include violent relaxation and two-body relaxation. Both of these
decrease initial structure, including ellipticity and substructure, although
some degree of structure is likely to remain long enough to be observable.
The ellipticity in the ONC could thus indicate an initially highly
aspherical initial condition.
Two-body relaxation also drives mass segregation although this occurs
over many dynamical times in large clusters such that the position of the
massive stars in the ONC reflects their initial location and thus constrains
how massive stars form.

Star-gas interactions include accretion of the gas onto the stars and
the feedback (especially from massive stars) from the stars onto the
gas.  Although feedback has not yet been studied in this context, we
are starting to understand the process of accretion in clusters. Gas
accretion in a stellar cluster is highly competitive and uneven. Stars
near the centre of the cluster accrete at significantly higher rates
due to their position where they are aided in attracting the gas by
the overall cluster potential. This competitive accretion naturally
results in both a spectrum of stellar masses, and an initial mass
segregation even if all the stars originate with equal
masses. Accretion in stellar clusters can also force the core of the
cluster to contract sufficiently to allow for stellar collisions to
occur. Such a collisional model for the formation of massive stars
evades the problem of accreting onto massive stars.

Wide binary systems in clusters are destroyed by stellar encounters in
clusters. The maximum separation which survives such interactions
depends primarily on the cluster density and velocity dispersion. Wide
systems are more likely to survive less dense clusters than in the
core of dense clusters. Binaries can thus be used as a tracer of the
cluster evolution.

Finally, clusters dissolve through either gas removal (generally
larger clusters) or through dynamical interactions which transfer all
the clusters binding enery to a central binary (small-N
clusters). Thus, clusters surviving to the main sequence and Galactic
cluster status represent a small subset of the initial population of
stellar clusters.

\end{document}